\documentclass[
  aps,
  prd,
  reprint,
  superscriptaddress,
  nofootinbib
]{revtex4-2}

\usepackage[T1]{fontenc}
\usepackage[utf8]{inputenc}
\usepackage{microtype}

\usepackage{amsmath,amssymb,bm}
\usepackage{slashed}
\usepackage{graphicx}
\usepackage{placeins}

\usepackage[dvipsnames]{xcolor}
\usepackage{tikz}
\usetikzlibrary{
  decorations.markings,
  decorations.pathmorphing
}

\usepackage{booktabs}
\usepackage{array}

\usepackage{hyperref}

\hypersetup{
  colorlinks=true,
  linkcolor=MidnightBlue,
  citecolor=MidnightBlue,
  urlcolor=MidnightBlue,
  pdftitle={Electron and Muon g-2 Constraints on Light Vector Bosons:
  Dark Photons and the $X_{17}$ Boson},
  pdfauthor={R. Serao and A. Capolupo}
}

\newcommand{\amu}{a_{\mu}}
\renewcommand{\ae}{a_{e}}
\newcommand{\eps}{\varepsilon}

\newcommand{\mAp}{m_{A'}}
\newcommand{\Ap}{A'}

\begin{document}

\title{Electron and Muon $g-2$ Constraints on Light Vector Bosons:
Dark Photons and the $X_{17}$ Boson}

\author{Raoul Serao}
\email{rserao@unisa.it}
\affiliation{
Department of Physics, Faculty of Science, University of Zagreb,
Bijeni\v{c}ka cesta 32, 10000 Zagreb, Croatia
}

\author{Antonio Capolupo}
\email{capolupo@sa.infn.it}
\affiliation{
Dipartimento di Fisica ``E.~R. Caianiello'',
Universit\`a di Salerno,
Via Giovanni Paolo II 132, 84084 Fisciano, Italy
}
\affiliation{
INFN -- Gruppo Collegato di Salerno,
Via Giovanni Paolo II 132, 84084 Fisciano, Italy
}

\begin{abstract}
In light of the latest experimental results from Fermilab on the muon anomalous magnetic moment, it is necessary to reassess existing models that invoke light vector bosons as a possible explanation of the g-2 anomaly. In this work, we revisit the phenomenological implications of the $X_{17}$ boson and the dark photon by considering these updated experimental results.
We combine the current experimental muon $g-2$ world average, which incorporates the final Fermilab result, with the latest electron $g-2$ determinations based on cesium and rubidium measurements to set 95\% CL exclusion contours for a pure vector mediator coupled to leptons.
 We explicitly test the assumption that the electron and muon coupling magnitudes are equal by comparing this restricted case with the case of independent electron and muon couplings and quantify the impact on the allowed parameter space. In the minimal visible dark-photon model, both leptons constrain the same kinetic mixing and are analyzed through a combined $\chi^2$ analysis. We compare the resulting $g-2$ bounds with existing accelerator direct-search exclusions and model-dependent astrophysical and cosmological constraints. From the accelerator comparison, we identify a region in the $(m_{A'},|\epsilon|)$ parameter space near $17$~MeV, close to the reported $X_{17}$ mass, that remains allowed by the direct-search contours displayed here but is excluded by the cesium-based electron $g-2$ constraint. The rubidium-based fit does not exclude this interval.
For an $X_{17}$ boson with independent lepton couplings, we constrain the electron and muon couplings separately. Electron-only direct searches leave two disconnected allowed regions near the reported $X_{17}$ mass: a newly reopened low-coupling interval and a higher-coupling region above the NA64 excluded band. The cesium-based electron $g-2$ constraint closes the higher-coupling region, while the rubidium-based constraint reduces its extent; neither affects the newly reopened low-coupling interval. Using the current experimental muon $g-2$ world average, we obtain a new $g-2$-based exclusion region for the muon coupling, with no significant preference for a nonzero coupling.
\end{abstract}


\maketitle
\section{Introduction}
\label{sec:introduction}

Light neutral vectors are among the simplest extensions of the Standard Model, and their contribution to charged-lepton anomalous magnetic moments is well known.  The central issue addressed here is not the loop calculation itself, but its interpretation.  The same positive correction is often used to discuss models with different coupling structures, production mechanisms, and experimental constraints.  In particular, a minimal dark photon and a flavor-dependent $X_{17}$ boson cannot be treated as a single hypothesis merely because they generate the same type of loop contribution.  Any current analysis must also incorporate the final Fermilab muon measurement and the revised Standard Model inputs rather than carry forward preferred regions obtained in an earlier precision landscape.

The minimal dark photon is the standard vector-portal realization.  Its couplings to electrically charged Standard Model particles are controlled by one kinetic-mixing parameter, so electron and muon magnetic-moment constraints can be compared directly with visible accelerator searches in a common parameter space.  This framework and its phenomenology have been developed extensively in model studies and broad surveys~\cite{Holdom1986,Pospelov2009,Bjorken2009,Batell2009FixedTarget,Batell2009BFactories,Essig2013,Fabbrichesi2020,Graham2021}.

The proposed $X_{17}$ interpretation of the nuclear pair-creation anomalies has a different phenomenological structure.  It originated from the anomalous angular distribution of electron--positron pairs in an excited beryllium transition~\cite{Krasznahorkay2016}; structures later reported in helium and carbon motivated further nuclear, direct, and model-building studies~\cite{Krasznahorkay2021,Krasznahorkay2022,Kozaczuk2017,DelleRose2017,Alves2023,Capolupo2025X17Anomalies,SeraoReview2026}.  The evidence remains unsettled: MEG~II found no significant signal in an independent beryllium study~\cite{MEGII2025}, while PADME reported its largest deviation near the proposed mass but without discovery-level global significance~\cite{PADME2025}.  A recent reanalysis of SLAC E141 has also changed the direct-search picture in this mass region~\cite{E141update2026}.  Unlike the minimal dark photon, a protophobic $X_{17}$ construction is not governed by a single kinetic-mixing parameter: suppressing the proton coupling requires a specific relation between the up- and down-quark couplings, while the lepton couplings are not fixed by the same relation~\cite{Feng2016,Feng2017}. Electron and muon measurements therefore need not constrain the same parameter.

The precision context has changed at the same time.  Fermilab E989 has completed its measurement of the positive muon anomalous magnetic moment~\cite{FermilabFinal2025,FinalReport2026}.  The current Muon g-2 Theory Initiative reference prediction, based on a consolidated lattice-QCD treatment of hadronic vacuum polarization, lies close to the experimental average~\cite{WP2025}; the large positive discrepancy inferred from the previous White Paper prediction is therefore no longer the appropriate starting point for a current constraint analysis.  The electron magnetic moment provides an independent and more coupling-sensitive low-mass test, but its Standard Model comparison depends on an external determination of the fine-structure constant.  Although the five-loop quantum-electrodynamics calculations are now mutually consistent~\cite{Volkov2024Electron,Aoyama2025Electron}, the cesium and rubidium recoil determinations remain mutually inconsistent~\cite{Parker2018,Morel2020,Fan2023} and must be treated as separate inputs.

Here we use the updated measurements of the electron and muon anomalous magnetic moments to derive mass-dependent constraints on the light vector couplings to electrons and muons. We use the values of the fine-structure constant obtained from cesium and rubidium atom measurements~\cite{Hanneke2008,Fan2023,Altmannshofer2026} to constrain the coupling of light vector bosons to electrons. It is worth noting that the rubidium-based determination yields $\Delta a_e^{\rm Rb}>0$, whereas the cesium-based determination gives $\Delta a_e^{\rm Cs}<0$. The corresponding determinations of $\alpha$ are separated by approximately $5.5\sigma$. Consequently, we regard them as statistically inconsistent alternative inputs and do not combine the associated constraints.  In particular, we analyze the dark photon and $X_{17}$ boson scenarios.

In the minimal visible dark-photon model~\cite{Fabbrichesi2020,Graham2021}, the electron and muon anomalous magnetic moments constrain the same kinetic-mixing parameter $\epsilon$. We compare the resulting $(g-2)$ bounds with existing accelerator direct-search exclusions and model dependent astrophysical and cosmological constraints. We identify a region near $m_{A'}\simeq17$~MeV, close to the reported $X_{17}$ mass, that remains allowed by the accelerator direct-search contours considered here but is excluded by the cesium-based electron $g-2$ constraint. Including the muon measurement consistently in the dark-photon fit only marginally modifies this bound. The rubidium-based fit does not exclude this interval.

We then consider the $X_{17}$ boson allowing the electron and muon couplings to vary independently, and derive the corresponding exclusion regions in the $(m_X,|\epsilon_e|)$ and $(m_X,|\epsilon_\mu|)$ planes. For the electron coupling, the updated direct-search landscape leaves two disconnected allowed regions near the reported $X_{17}$ mass. The first is the newly reopened low coupling visible search window, which remains unaffected by the electron $g-2$ constraints. A second allowed region appears above the upper edge of the NA64 excluded band. This higher coupling region is completely removed by the Cs-based $g-2$ bound, whereas the Rb-based bound excludes only its upper part. For the muon coupling, we obtain a new $g-2$-based exclusion region using the final Fermilab result together with the current Standard Model prediction.

The light-vector interaction is introduced in Section~\ref{sec:framework}. Section~\ref{sec:g2} presents the one-loop contribution and the statistical analysis. The dark photon and $X_{17}$ results
are discussed in Section~\ref{sec:darkphoton}. The conclusions are drawn in Section~\ref{sec:conclusions}.

\section{Effective descriptions of light vector bosons}
\label{sec:framework}

For a generic neutral spin-one state $V_\mu$ we use
\begin{equation}
\mathcal L_V = -\frac14 V_{\mu\nu}V^{\mu\nu}
 +\frac12m_V^2V_\mu V^\mu
 +e\sum_f\eps_f V_\mu\bar f\gamma^\mu f,
\label{eq:genericL}
\end{equation}
where $\mathcal L_V$ is the effective Lagrangian density, $V_\mu$ is the vector field, $V_{\mu\nu}\equiv\partial_\mu V_\nu-\partial_\nu V_\mu$ is its field-strength tensor, and $m_V$ is its mass.  The index $f$ labels a Dirac-fermion species, and the sum runs over the fermions coupled to $V_\mu$.  The positive quantity $e$ is the electromagnetic coupling and $\alpha\equiv e^2/(4\pi)$ is the fine-structure constant.  The $\eps_f$ are signed dimensionless vector couplings normalized to $e$.  At nucleon level we write $\eps_p$ and $\eps_n$ for the corresponding effective proton and neutron couplings.
In the following we discuss the main  differences between dark photons and $X_{17}$.

- The Lagrangian for the minimal kinetically mixed dark photon is given by
\begin{equation}
\begin{aligned}
\mathcal L_{\rm kin}\supset{}&
 -\frac14 F_{\mu\nu}F^{\mu\nu}
 -\frac14 F'_{\mu\nu}F'^{\mu\nu}\\
&-\frac{\eps}{2}F_{\mu\nu}F'^{\mu\nu}
 +\frac12\mAp^2\Ap_\mu\Ap^\mu .
\end{aligned}
\label{eq:kinmix}
\end{equation}
Here $F_{\mu\nu}$ and $F'_{\mu\nu}$ are respectively the ordinary-photon and dark-photon field-strength tensors, $\mAp$ is the dark photon mass, and $\eps$ is the dimensionless kinetic-mixing parameter.
After diagonalizing the kinetic terms and neglecting electroweak-suppressed corrections, we obtain the dark photon interaction Lagrangian density
\begin{equation}
\mathcal L_{\Ap,\rm int}=\eps e\,\Ap_\mu J_{\rm EM}^\mu,
\qquad \eps_f=\eps Q_f.
\label{eq:dpmap}
\end{equation}
Here $J_{\rm EM}^\mu\equiv\sum_f Q_f\bar f\gamma^\mu f$ is the electromagnetic current, and $Q_f$ is the electric charge of fermion $f$ in units of $e$.  Thus $|\eps_e|=|\eps_\mu|=|\eps|$ and $\eps_p=+\eps$, with the fermion charge sign carried by $Q_f$.  This correlation allows us to obtain an exclusion plot in $(\mAp,\eps)$ without making an additional flavor hypothesis.

Notice that electron beam dumps experiments constrain $\eps_e$, whereas the muon anomalous magnetic moment constrains $\eps_\mu$. 
We distinguish the minimal \emph{visible} dark photon, with no lighter
dark-sector states and therefore no invisible decay channels, from models in
which $\Ap$ can decay predominantly into lighter dark-sector states.

- For the $X_{17}$ boson, by contrast, the couplings $\epsilon_f$ are not fixed by a single kinetic-mixing parameter. We therefore treat $\epsilon_e$ and $\epsilon_\mu$ as independent parameters and use $m_X=16.88\pm0.05$~MeV~\cite{Combined2026}.
The comparison with visible electron searches assumes negligible invisible
and neutrino couplings and $\operatorname{BR}(X\to e^+e^-)=1$.
Only searches whose production rate is fixed by the electron coupling are transferred directly to the $(m_{X},|\eps_e|)$ plane. Limits based on meson decays, proton bremsstrahlung or other hadronic production channels require an explicit framework for the quark or nucleon, couplings and are not applied here. For PADME, which define $\mathcal{L}\supset g_{v e} X_{17}^\mu \bar{e}\gamma_\mu e$, we use $|\eps_e|=|g_{v e}/e$.

\section{Lepton \texorpdfstring{$g-2$}{g-2} after the final Fermilab result}
\label{sec:g2}
Building on Ref.~\cite{Capolupo2025X17Anomalies}, we compute the one-loop contribution of a light vector boson to the electron and muon anomalous magnetic moments. We then use the current (g-2) data to constrain the corresponding couplings in the $X_{17}$ and dark photon scenarios.
For a charged lepton $\ell\in\{e,\mu\}$ with gyromagnetic factor $g_\ell$, the magnetic anomaly and its experiment--Standard Model residual are defined as
$a_\ell\equiv\frac{g_\ell-2}{2},$$\Delta a_\ell\equiv a_\ell^{\rm exp}-a_\ell^{\rm SM}$.
The superscripts ``exp'' and ``SM'' identify, respectively, the experimentally measured value and the corresponding Standard Model prediction.

\begin{figure}[t]
\centering
\begin{tikzpicture}[scale=1.05,every node/.style={font=\small}]
\tikzset{
  fermion/.style={thick,postaction={decorate},decoration={markings,mark=at position 0.58 with {\arrow{>}}}},
  boson/.style={thick,decorate,decoration={snake,amplitude=1.1pt,segment length=5.2pt}}
}
\coordinate (pin) at (-1.75,-1.15);
\coordinate (v1) at (-0.72,-0.56);
\coordinate (vc) at (0.30,0);
\coordinate (v2) at (-0.72,0.56);
\coordinate (pout) at (-1.75,1.15);
\coordinate (photon) at (2.05,0);
\draw[fermion] (pin) -- (v1) -- (vc);
\draw[fermion] (vc) -- (v2) -- (pout);
\draw[boson] (v1) -- node[left=2pt] {$V(k)$} (v2);
\draw[boson] (vc) -- node[above=3pt] {$\gamma(q)$} (photon);
\fill (v1) circle (1.0pt);
\fill (vc) circle (1.0pt);
\fill (v2) circle (1.0pt);
\node[below left=1pt] at (pin) {$\ell(p)$};
\node[above left=1pt] at (pout) {$\ell(p')$};
\end{tikzpicture}
\caption{One-loop light-vector correction to the on-shell lepton--photon vertex.  The internal wavy line denotes the vector $V$, while the external wavy line is the photon.  The momentum routing is defined in the text; compare Ref.~\cite{Capolupo2025X17Anomalies}.}
\label{fig:g2vertex}
\end{figure}
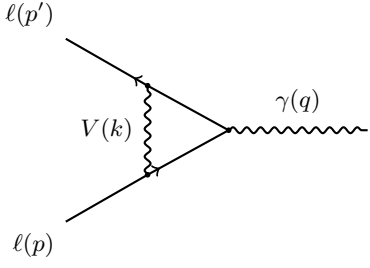

For the interaction in Eq.~\eqref{eq:genericL}, the relevant one-loop correction, induced by the presence of a vector field, is the modification of the on-shell lepton--photon vertex shown in Fig.~\ref{fig:g2vertex}.  The calculation presented here for a generic vector mass $m_V$ and coupling $\eps_\ell$ follows the $X_{17}$ derivation of Ref.~\cite{Capolupo2025X17Anomalies}.

With $p^2=p'^2=m_\ell^2$ and $q=p'-p$, and after factoring out the external electromagnetic coupling, the vector contribution to the vertex in Feynman gauge can be written as
\begin{equation}
\begin{aligned}
\delta\Gamma_V^\mu(p',p)
={}&-i e^2\eps_\ell^2\!\int\!\frac{d^4k}{(2\pi)^4}
\frac{1}{D_{p'}D_pD_V}\\
&\times\gamma^\alpha(\slashed p'-\slashed k+m_\ell)\gamma^\mu \times(\slashed p-\slashed k+m_\ell)\gamma_\alpha .
\end{aligned}
\label{eq:vertexloop}
\end{equation}
where $D_{p'}=(p'-k)^2-m_\ell^2+i\eta$, $D_p=(p-k)^2-m_\ell^2+i\eta$, and $D_V=k^2-m_V^2+i\eta $.  Gauge-dependent longitudinal terms do not contribute to the on-shell Pauli form factor for the conserved vector current.  The renormalized vertex between on-shell spinors is decomposed as
\begin{equation}
\Gamma^\mu(q)=F_1(q^2)\gamma^\mu+
\frac{i\sigma^{\mu\nu}q_\nu}{2m_\ell}F_2(q^2),
\qquad
\sigma^{\mu\nu}\equiv\frac{i}{2}[\gamma^\mu,\gamma^\nu],
\label{eq:vertexdecomposition}
\end{equation}
where $F_1(q^2)$ and $F_2(q^2)$ are form factors \cite{Schwartz:2014sze}.
The vector induced magnetic anomaly coincides with the Pauli form factor at zero momentum transfer, $\Delta a_\ell^V=F_2^V(0)$.  Combining the three propagator denominators, shifting the loop momentum, and retaining the coefficient of $i\sigma^{\mu\nu}q_\nu$ gives
\begin{equation}
F_2^V(0)=\frac{e^2\eps_\ell^2}{8\pi^2}
\int_0^1 dz\,
\frac{2z(1-z)^2m_\ell^2}
{(1-z)^2m_\ell^2+z m_V^2}.
\label{eq:pauliformfactor}
\end{equation}
Using $\alpha=e^2/(4\pi)$ and defining $r\equiv m_V/m_\ell$, Eq.~\eqref{eq:pauliformfactor} becomes
\begin{align}
\Delta a_\ell^V(m_V)
 &= \frac{\alpha\eps_\ell^2}{2\pi}F_V\!\left(\frac{m_V}{m_\ell}\right),
\label{eq:g2master}\\
F_V(r)&=\int_0^1 dz\,
\frac{2z(1-z)^2}{(1-z)^2+r^2z}.
\label{eq:loopfunction}
\end{align}
Here $z$ is a Feynman parameter and $F_V(r)$ is the dimensionless vector loop function.  The shift $\Delta a_\ell^V$ is positive for a real vector coupling.  The familiar limits $F_V(0)=1$ and $F_V(r\gg1)\simeq2/(3r^2)$ provide useful checks on the numerical evaluation.
We treat the experimental and Standard Model uncertainties as Gaussian and combine them in quadrature in $\sigma_{\Delta a_\ell}$. The electron and muon residuals are treated as statistically independent.
For convenience, we introduce the non-negative parameter $q_l\equiv\eps_l^2$, where $\eps_\ell$ denotes the coupling of the vector boson to the lepton $\ell$
and perform the $\chi^2$ analysis directly in terms of $q_\ell$.
  At fixed $m_V$,
\begin{align}
\chi^2(q_\ell;m_V)&=
\frac{[\Delta a_\ell-C_\ell(m_V)q_\ell]^2}{\sigma_{\Delta a_{\ell}}^2},
\label{eq:chi2}\\
C_\ell(m_V)&=\frac{\alpha}{2\pi}F_V(m_V/m_\ell),
\label{eq:Cmu}
\end{align}
where $\chi^2$ is the Gaussian goodness-of-fit statistic, $\Delta a_\ell\equiv a_\ell^{\rm exp}-a_\ell^{\rm SM}$, $C_\ell(m_V)$ is the coefficient converting $q_\ell$ into the predicted shift $\Delta a_\ell^V=C_\ell q_\ell$, and $\sigma_{\Delta a_{\ell}}$ is the one-standard-deviation uncertainty of $\Delta a_l$. The physical best fit is
\begin{equation}
\widehat q_\ell=
\max\left(0,\frac{\Delta a_\ell}{C_\ell}\right).
\label{eq:physicalbest}
\end{equation}
We define the one-sided 95\% upper bound by
\begin{equation}
\chi_\ell^2(q_{\ell,95})-
\chi_\ell^2(\widehat q_\ell)=z_{0.95}^2,
\qquad z_{0.95}=1.645.
\label{eq:deltachi2}
\end{equation}
Solving Eq.~\eqref{eq:deltachi2} subject to the physical constraint
$q_\ell\geq0$ gives
\begin{equation}
\begin{aligned}
q_{\ell,95}
&=
\frac{1}{C_\ell}
\begin{cases}
\Delta a_\ell+z_{0.95}\sigma_{\Delta a_\ell},
& \Delta a_\ell\geq 0,\\[2mm]
\Delta a_\ell+
\sqrt{(\Delta a_\ell)^2+
z_{0.95}^2\sigma_{\Delta a_\ell}^2},
& \Delta a_\ell<0,
\end{cases}
\end{aligned}
\label{eq:upper}
\end{equation}
then we define $|\eps_\ell|_{95}=\sqrt{q_{\ell,95}}.$
The second branch is required for the negative cesium-based electron
residual.

For the  $X_{17}$ boson, Eq.~\eqref{eq:upper} is applied independently
to the electron and muon couplings. In the minimal dark-photon model, both
leptons instead constrain the same parameter $q=\eps^2$. For each electron
input $\Gamma\in\{\mathrm{Cs},\mathrm{Rb}\}$, we therefore define
\begin{equation}
\chi_{\mathrm{DP},\Gamma}^2(q;m_V)=
\chi_\mu^2(q;m_V)+\chi_{e,\Gamma}^2(q;m_V).
\label{eq:jointchi2}
\end{equation}
Introducing
\begin{align}
A_\Gamma &=
\frac{C_\mu^2}{\sigma_{\Delta a_\mu}^2}
+\frac{C_e^2}{(\sigma_{\Delta a_e}^{\Gamma})^2},
\nonumber\\
B_\Gamma &=
\frac{C_\mu\Delta a_\mu}{\sigma_{\Delta a_\mu}^2}
+\frac{C_e\Delta a_e^{\Gamma}}
{(\sigma_{\Delta a_e}^{\Gamma})^2}.
\label{eq:jointAB}
\end{align}
the physical best fit is
$\widehat q_\Gamma=\max(0,B_\Gamma/A_\Gamma)$, while the one-sided upper bound is
\begin{equation}
\begin{aligned}
q_{95,\Gamma}
&=
\begin{cases}
\dfrac{B_\Gamma}{A_\Gamma}
+\dfrac{z_{0.95}}{\sqrt{A_\Gamma}},
& B_\Gamma\geq 0,\\[3mm]
\dfrac{B_\Gamma+
\sqrt{B_\Gamma^2+z_{0.95}^2A_\Gamma}}{A_\Gamma},
& B_\Gamma<0,
\end{cases}
\end{aligned}
\label{eq:jointupper}
\end{equation}
and we define $|\eps|_{95,\Gamma}=\sqrt{q_{95,\Gamma}}.$
The cesium and rubidium determinations are treated as alternative electron
inputs and are never combined with each other.

For the muon, we use the current experimental world average, which incorporates the final Fermilab result, and the current Standard Model prediction:
 $\amu^{\rm exp}=116592071.5(14.5)\times10^{-11},$ ~\cite{FinalReport2026}
and $\amu^{\rm SM}=116592033(62)\times10^{-11}$~\cite{WP2025} which gives $\Delta\amu\pm \sigma_{\Delta a_{\mu}}=(38.5\pm63.7)\times10^{-11}.$
For the electron, the corresponding residual are:
$\Delta\ae^{\rm Rb}\pm\sigma_{\Delta a_{e}}^{\rm Rb} =(35\pm16)\times10^{-14},\quad$$
\Delta\ae^{\rm Cs}\pm\sigma_{\Delta a_{e}}^{\rm Cs}  =(-100\pm26)\times10^{-14}$, as derived for cesium and rubidium ~\cite{Hanneke2008,Fan2023,Altmannshofer2026}.

\section{$g-2$ constraints on the minimal dark photon and $X_{17}$ boson}
\label{sec:darkphoton}

\begin{figure*}[t]
\centering
\includegraphics[width=0.8\textwidth]{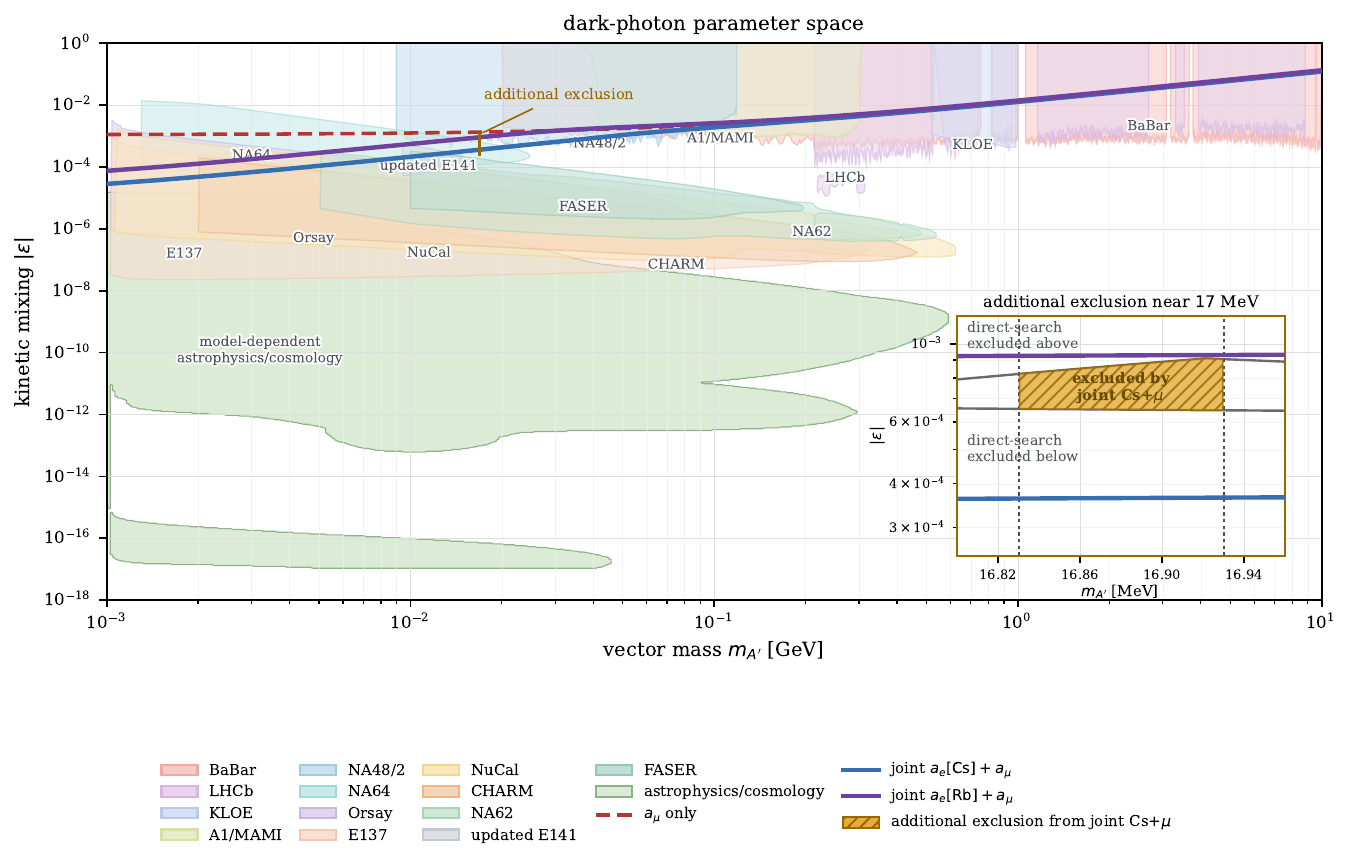}
\caption{Constraints on the minimal visible dark photon. The pale-green
regions show the model-dependent astrophysical and cosmological exclusions
of Refs.~\cite{CaputoHeavyDP2025,AxionLimits}, while the distinct pastel
regions identify the individual accelerator direct-search exclusions. The
FASER 2026 preliminary contour~\cite{FASER2026} and the revised E141
contour~\cite{E141update2026} are shown at 90\% C.L. The dashed red curve denotes the muon-only 95\% upper bound, while the
solid blue and violet curves denote the joint Cs+$\mu$ and Rb+$\mu$
95\% upper bounds, respectively. The inset highlights the additional
excluded region derived in this work near the reported $X_{17}$ mass: this
region is allowed by the accelerator direct searches shown here but is
excluded by the joint Cs+$\mu$ bound. The Cs and Rb determinations are treated as
alternative electron inputs and are not combined with each other. }
\label{fig:darkphoton}
\end{figure*}

\textit{Dark photon: -} For the minimal dark-photon model, Eq.~\eqref{eq:dpmap} implies $|\epsilon_e|=|\eps_\mu|=|\eps|$.
For each electron input $\Gamma\in\{\mathrm{Cs},\mathrm{Rb}\}$, we use the joint statistic in Eq.~\eqref{eq:jointchi2} and the upper bound in
Eq.~\eqref{eq:jointupper} to derive $|\eps|_{95,\Gamma}$ as a function of the
dark-photon mass $m_{A'}$. In Fig.~\ref{fig:darkphoton}, we compare the muon-only and joint
Cs+$\mu$ and Rb+$\mu$ 95\% upper bounds on $|\eps|$ with the accelerator
direct-search exclusions and the model dependent astrophysical and
cosmological constraints~\cite{FASER2026,Bjorken2009,Blumlein2014,
Andreas2012,NA622024,NA482015,LHCb2020,NA642020,BaBar2014,
A1MAMI2014,KLOE2013,KLOE2018,E141update2026,CaputoHeavyDP2025,
AxionLimits}. The inset highlights the additional excluded region derived
in this work near the reported $X_{17}$ mass: this region remains allowed by
the accelerator direct searches alone but is excluded by our result on the joint
Cs+$\mu$ $g-2$ bound.
We include in Fig.~\ref{fig:darkphoton} the revised E141 exclusion contour obtained in the
2026 reanalysis~\cite{E141update2026}, replacing the legacy E141 recast
commonly shown in earlier dark-photon compilations. Under the minimal
visible dark-photon assumptions adopted here, the production and visible
decay rates relevant for an electron beam-dump experiment are controlled
by the same kinetic-mixing parameter $\epsilon$, allowing the E141 result
to be represented directly in the $(m_{A'},|\epsilon|)$ plane. The revised
analysis substantially weakens the E141 sensitivity in the
$m_{A'}\sim 14$--$18\,\mathrm{MeV}$ region and no longer excludes
$m_{A'}=16.88\,\mathrm{MeV}$.

At $m_{A'}=16.88$~MeV, the physical best fits and one-sided 95\% upper
bounds are
$
|\widehat{\eps}|_{\mathrm{Cs}}=0,\quad
|\eps|_{95,\mathrm{Cs}}=3.64\times10^{-4},
\quad
|\widehat{\eps}|_{\mathrm{Rb}}=7.06\times10^{-4},
\quad
|\eps|_{95,\mathrm{Rb}}=9.28\times10^{-4}.$
At the same mass, the accelerator direct-search exclusions leave the approximate interval
$ 6.5\times10^{-4}<|\eps|<8.7\times10^{-4}.$
This interval is obtained from the displayed exclusion contours and should
not be interpreted as the result of a combined experimental likelihood.
The joint Cs+$\mu$ $g-2$ bound excludes this remaining interval, whereas
the corresponding Rb+$\mu$ bound does not.

\begin{figure*}[t]
\centering
\includegraphics[width=0.8\textwidth]{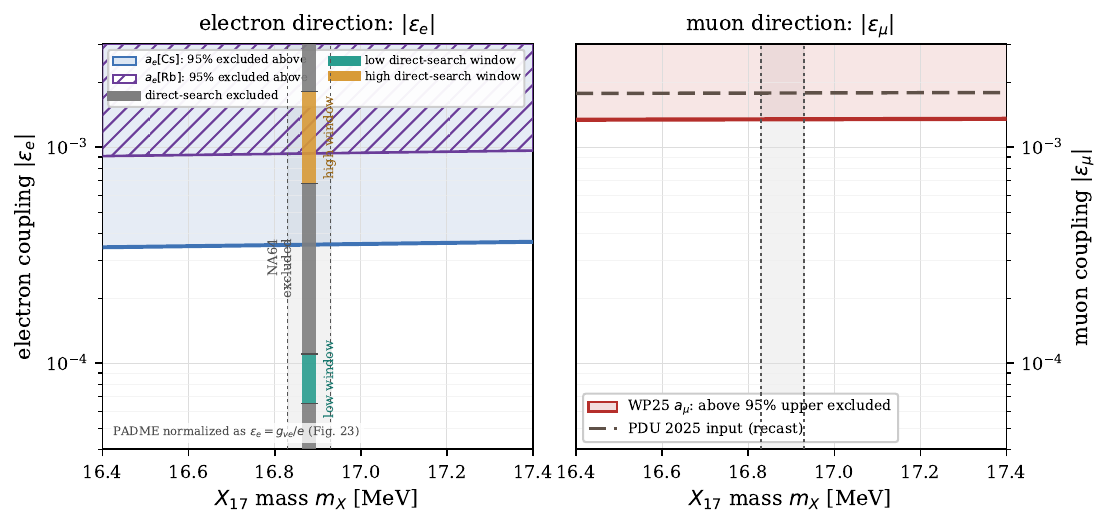}
\caption{Excluded regions for the independent $X_{17}$ electron and
muon couplings. The left panel shows the
solid-blue Cs-based and dashed-violet Rb-based electron bounds.
At the reference mass, gray segments denote electron only direct search exclusions, while teal and amber denote the low and high coupling surviving windows, respectively. The NA64 excluded band lies between the two windows. The upper edge of the amber window is the PADME observed 90\% C.L. upper limit after converting its coupling convention with $\eps_e=g_{v e}/e$. These reference mass segments are an overlay of published and graphically extracted boundaries. Hadronic FASER and NA48/2 dark photon limits are not applied to the independent $\eps_e$ axis.
The right panel shows the muon upper bound. Couplings above each curve are excluded
at one-sided 95\% C.L. The vertical gray bands mark
$m_X=16.88\pm0.05$~MeV. The electron and muon couplings are treated
independently, and the Cs and Rb results are alternative electron inputs.}
\label{fig:x17}
\end{figure*}

\textit{- $X_{17}$ boson: -}
We now consider the $X_{17}$ boson at the reference mass
$m_X=16.88$~MeV~\cite{Combined2026}. For the muon, Eq.~\eqref{eq:loopfunction} gives
$ F_V\left(\frac{16.88}{105.658}\right)=0.68017,\quad 
C_\mu=7.8996\times10^{-4}.$
Using the muon residual given in Sec.~\ref{sec:g2}, the physical best fit
in Eq.~\eqref{eq:physicalbest} and the upper bound in
Eq.~\eqref{eq:upper} give
$ |\widehat{\eps}_\mu|=6.98\times10^{-4},
\qquad
|\eps_\mu|_{95}=1.35\times10^{-3}.$
The nonzero best fit reflects the positive central value of
$\Delta a_\mu$, but its significance is only $0.60\sigma$ and therefore
does not provide evidence for a nonzero muon coupling.
For the electron, Eq.~\eqref{eq:loopfunction} gives
$ F_V\left(\frac{16.88}{0.510999}\right) =6.0265\times10^{-4},
\quad
C_e=6.9992\times10^{-7}.$
Using the Rb- and Cs-based residuals, the
physical best fits and upper bounds are
$|\widehat{\eps}_e|^{\mathrm{Rb}}
=7.07\times10^{-4},
\quad |\eps_e|_{95}^{\mathrm{Rb}}=9.36\times10^{-4},
\quad
|\widehat{\eps}_e|^{\mathrm{Cs}}=0,
\quad
|\eps_e|_{95}^{\mathrm{Cs}}
=3.54\times10^{-4}.$
The Rb and Cs results are alternative constraints and are not combined.
Under the visible-vector assumptions specified in Sec.~\ref{sec:framework}, electron-only direct searches at $m_X=16.88$~MeV leave two approximately allowed intervals. The newly reopened low-coupling window is $6.5\times10^{-5}\lesssim |\eps_e| \lesssim 1.1\times10^{-4},$
bounded approximately by the Orsay recast from below and by the lower edge of the NA64 excluded band from above. At this mass, NA64 excludes approximately
$1.1\times10^{-4}\lesssim |\eps_e| \lesssim 6.5\times10^{-4}$ at 90\% C.L. A second allowed region therefore opens above the upper edge of the NA64 exclusion and extends up to the PADME observed upper limit. From the PADME contour, we extract approximately $g_{ve}\simeq5.48\times10^{-4}$, which in our coupling convention, $g_{ve}=e\eps_e$, corresponds to $|\eps_e|\simeq \frac{g_{ve}}{e}\simeq1.81\times10^{-3}.$
The Cs-based electron $g-2$ bound excludes this higher-coupling region completely, whereas the Rb-based bound reduces it to approximately $6.5\times10^{-4}\lesssim |\eps_e| \lesssim 9.36\times10^{-4}.$
Both electron $g-2$ constraints leave the newly reopened low-coupling window unchanged.
Figure~\ref{fig:x17} shows the corresponding 95\% excluded regions in
the $(m_X,|\eps_e|)$ and $(m_X,|\eps_\mu|)$ planes. The left panel
compares the electron $g-2$ upper bounds with the visible direct-search
exclusion, while the right panel shows the muon $g-2$ exclusion.
At the reference mass, imposing the relation $q_e=q_\mu$
increases the minimum $\chi^2$ relative to the independent fit by
$2.3\times10^{-4}$ for the Rb input and by $0.37$ for the Cs input.The current $g-2$ data alone therefore do not significantly distinguish the hypothesis of equal electron and muon coupling magnitudes from the hypothesis in which the two coupling magnitudes are independent. Their different phenomenological implications arise from how the electron and muon couplings are related to
the direct-search constraints.

\section{Conclusions}
\label{sec:conclusions}

The final Fermilab measurement and the WP2025 Standard Model prediction have changed the interpretation of the constraints on light vector bosons. In this work, we used the updated electron and muon anomalous magnetic moments to derive mass-dependent 95\% upper bounds on the couplings of light vector bosons to electrons and muons. We then applied these bounds to the minimal visible dark photon and to the $X_{17}$ boson.

The electron constraints were obtained using the independent cesium- and rubidium-based determinations of the fine-structure constant~\cite{Parker2018,Morel2020}. The rubidium-based analysis yields a positive value of $\Delta a_e^{\rm Rb}$, whereas the cesium-based analysis yields a negative value of $\Delta a_e^{\rm Cs}$. The corresponding determinations of $\alpha$ differ by approximately $5.5\sigma$. We therefore treated them as statistically incompatible alternative inputs and did not combine the resulting constraints. A more detailed analysis of this issue will be presented elsewhere.

In the minimal visible dark-photon model~\cite{Fabbrichesi2020,Graham2021},
the electron and muon anomalous magnetic moments constrain the same
kinetic-mixing parameter $\epsilon$. We combined the two measurements and
compared the resulting $g-2$ upper bounds with the accelerator
direct-search exclusions and the model-dependent astrophysical and
cosmological constraints. Relative to the accelerator contours, the
Cs-based joint bound excludes an additional region near
$m_{A'}\simeq17$~MeV, close to the reported $X_{17}$ mass, whereas the
Rb-based joint bound does not exclude this interval.
For the $X_{17}$ boson, we treated the electron and muon couplings as independent parameters. In the $(m_X,|\epsilon_e|)$ plane, we compared the electron $g-2$ bounds with the visible direct-search exclusions~\cite{Andreas2012,NA642020,KLOE2013,PADME2025,E141update2026}. Near the reported $X_{17}$ mass, electron only direct searches leave two disconnected allowed regions. The newly reopened low coupling window remains unaffected by both electron $g-2$ constraints. By contrast, the second allowed region above the NA64 excluded band is completely excluded by the Cs-based bound and partially constrained by the Rb-based bound, which leaves couplings up to $|\epsilon_e|=9.36\times10^{-4}$ allowed. In the $(m_X,|\epsilon_\mu|)$ plane, we obtained a new $g-2$-based mass-dependent upper bound using the final Fermilab result and the current Standard Model prediction.
These results show that the dark photon and the $X_{17}$ boson must be treated as different coupling hypotheses. The minimal dark photon is constrained by a single kinetic-mixing parameter, whereas the $X_{17}$ scenario contains independent electron and muon couplings. Although the current $g-2$ data alone do not significantly distinguish the fit with equal electron and muon coupling magnitudes from the fit in which the two coupling magnitudes are independent, the two coupling hypotheses lead to different phenomenological interpretations when direct-search constraints are included.
 The updated inputs provide no robust evidence for a nonzero
light-vector coupling and primarily determine upper bounds.
\section*{Acknowledgements}
A.C.  acknowledges partial support from MIUR and INFN. A.C. also
acknowledges COST Action CA1511, Cosmology and Astrophysics Network for
Theoretical Advances and Training Actions (CANTATA).

\bibliographystyle{apsrev4-2}
\bibliography{references}

\end{document}